\documentstyle[manuscript,aps,epsfig]{revtex}
\begin{document}

% \preprint{WUHEP-00-18}
% \tighten

\title{Quantum Complex H\'enon-Heiles Potentials}

\author{Carl M. Bender$^1$, Gerald V. Dunne$^2$, Peter N. Meisinger$^1$, and
Mehmet \d{S}im\d{s}ek$^1$\cite{bye}}

\address{${}^1$Department of Physics, Washington University, St. Louis,
MO 63130, USA}

\address{${}^2$Department of Physics, University of Connecticut, Storrs, CT
06269, USA}

\date{\today}

\maketitle

\begin{abstract}
Quantum-mechanical ${\cal PT}$-symmetric theories associated with complex cubic
potentials such as $V=x^2+y^2+igxy^2$ and $V=x^2+y^2+z^2+igxyz$, where $g$ is a
real parameter, are investigated. These theories appear to possess real,
positive spectra. Low-lying energy levels are calculated to very high order in
perturbation theory. The large-order behavior of the perturbation coefficients
is determined using multidimensional WKB tunneling techniques. This approach
is also applied to the complex H\'enon-Heiles potential $V=x^2+y^2+ig(xy^2-{1
\over3}x^3)$.
\end{abstract}

% \pacs{PACS number(s): 02.30.Mv, 11.10.Kk, 11.10.Lm, 11.30.Er}

\vspace{.3in}
In this Letter we examine complex ${\cal PT}$-symmetric cubic Hamiltonians such
as
\begin{eqnarray}
H^{(2)} &\equiv& p_x^2+p_y^2+x^2+y^2+igxy^2,\nonumber\\
H^{(3)} &\equiv& p_x^2+p_y^2+p_z^2+x^2+y^2+z^2+igxyz,
\label{e1}
\end{eqnarray}
where $g$ is a real parameter \cite{BB}. The superscript on $H$ indicates the
number of degrees of freedom; these Hamiltonians are several-degree-of-freedom
generalizations of the one-dimensional complex cubic Hamiltonian $H^{(1)}=p^2+
x^2+igx^3$, which has recently been studied in great detail by many authors
\cite{CU1,CU2,CU3,CU4,LO1,CU5,CU6,CU7,CU8,CU9,LO2}. The non-Hermitian
Hamiltonian $H^{(1)}$ is interesting because its spectrum is entirely real and
positive. The reality of the spectrum is apparently due to the ${\cal PT}$
invariance of the Hamiltonian.

While many different one-degree-of-freedom examples of non-Hermitian ${\cal
PT}$-symmetric quantum systems have been studied, no multidimensional complex
${\cal PT}$-symmetric coupled-oscillator systems have been examined. The purpose
of this Letter is to show that (i) the property of real, positive spectra
persists even for quantum systems having several degrees of freedom, and (ii)
these theories have many other properties in common with theories described by
conventional Hermitian Hamiltonians.

Direct numerical evidence for the reality and positivity of the spectrum of
$H^{(1)}$ can be found by performing a Runge-Kutta integration of the associated
complex Schr\"odinger equation \cite{CU1}. Alternatively, the large-energy
eigenvalues of the spectrum can be calculated with great accuracy by using
conventional WKB techniques \cite{BO}. A strong argument for the reality and
positivity of the spectrum can be obtained by calculating the spectral zeta
function $Z(1)$ (the sum of the inverses of the eigenvalues). For the
Hamiltonian\footnote{This Hamiltonian, the {\it massless} case of $H^{(1)}$, has
a positive discrete spectrum. It is not known if the massless versions of
$H^{(2)}$ and $H^{(3)}$ have discrete spectra. Indeed, even for the massless
coupled anharmonic oscillator potential $V=x^2y^2$, it is also not known if the
spectrum is discrete.} $H=p^2+ix^3$ this was done by Mezincescu \cite{CU6} and
Bender and Wang \cite{CU8}. The exact result for $Z(1)$ is
\begin{equation}
Z(1)={4\sin^2(\pi/5)\Gamma^2(1/5)\over5^{6/5}\Gamma(3/5)}.
\label{e2}
\end{equation}
Using the numerical values of the first few eigenvalues and the WKB formula for
the high eigenvalues, one can conclude that any complex eigenvalues must be
larger in magnitude than about $10^{18}$. Some rigorous results regarding the
reality of the eigenvalues of $H^{(1)}$ have been obtained by Shin \cite{CU9},
who showed that the entire spectrum must lie in a narrow wedge containing the
positive-real axis. Other results have been obtained by Delabaere {\it et al}
\cite{CU2,CU5}.

Let us now return to the Hamiltonians in (\ref{e1}). The Schr\"odinger
equations associated with $H^{(1)}$, $H^{(2)}$, and $H^{(3)}$ are
\begin{eqnarray}
-\psi_1''(x)+(x^2+igx^3)\psi_1(x)&=&E\psi_1(x),\nonumber\\
-\nabla^2\psi_2(x,y)+(x^2+y^2+igxy^2)\psi_2(x,y)&=&E\psi_2(x,y),\nonumber\\
-\nabla^2\psi_3(x,y,z)+(x^2+y^2+z^2+igxyz)\psi_3(x,y,z)&=&E\psi_3(x,y,z).
\label{e3}
\end{eqnarray}
We have solved the Schr\"odinger equations (\ref{e3}) for the eigenvalues in
several ways. One technique is to diagonalize each Hamiltonian in a set of
multidimensional harmonic oscillator basis states. This procedure immediately
reveals that the energy levels are real.

A more precise calculation of the energies of the complex ${\cal PT}$-symmetric
Hamiltonians in (\ref{e1}) is performed using high-order Rayleigh-Schr\"odinger
perturbation theory. This technique was used in Refs.~\cite{LO1} and \cite{LO2}
to obtain the perturbation series for the ground-state energy of $H^{(1)}$. In
these references it was found that the Rayleigh-Schr\"odinger perturbation
series is Borel summable and that Pad\'e summation is in excellent agreement
with the real energy spectrum. Furthermore, Pad\'e analysis provides strong
numerical evidence that the once-subtracted ground-state energy considered as a
function of $g^2$ is a Stieltjes function.

\begin{figure}[h]
\centering{\epsfig{file=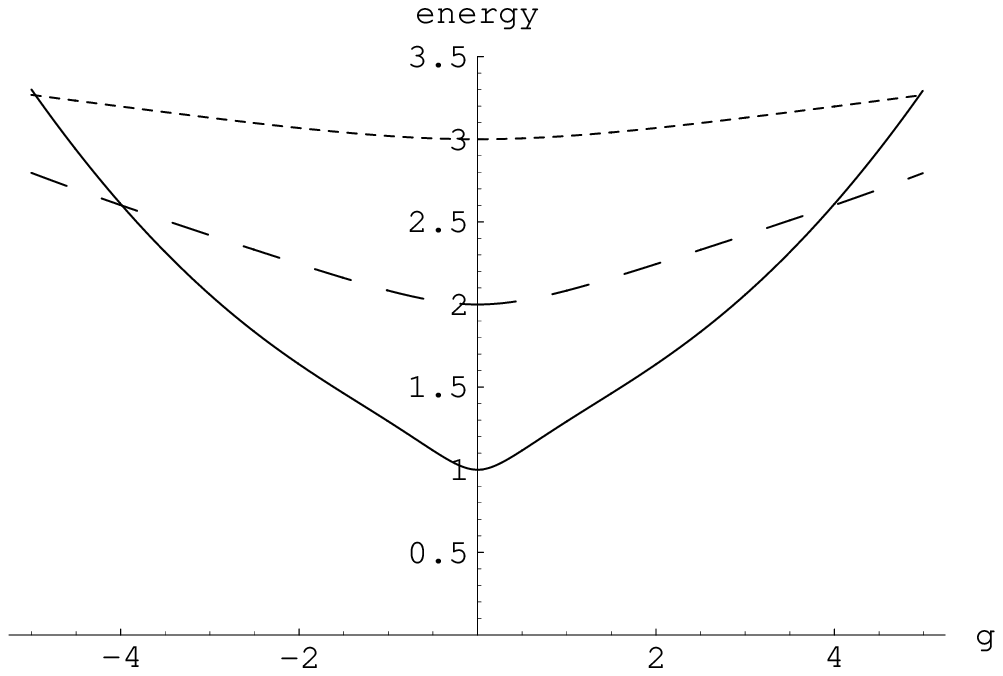}}
\vspace{1cm}
\caption{Ground-state energies of the Hamiltonians $H^{(1)}$ (solid line),
$H^{(2)}$ (long-dashed line), and $H^{(3)}$ (short-dashed line), as
functions
of the coupling constant $g$. Note that the energy levels are real and
positive. The graphs were obtained from the $(9,9)$ Pad\'e constructed from
the once-subtracted perturbation series for these energy levels.}
\label{f1}
\end{figure}

\begin{figure}
\centering{\epsfig{file=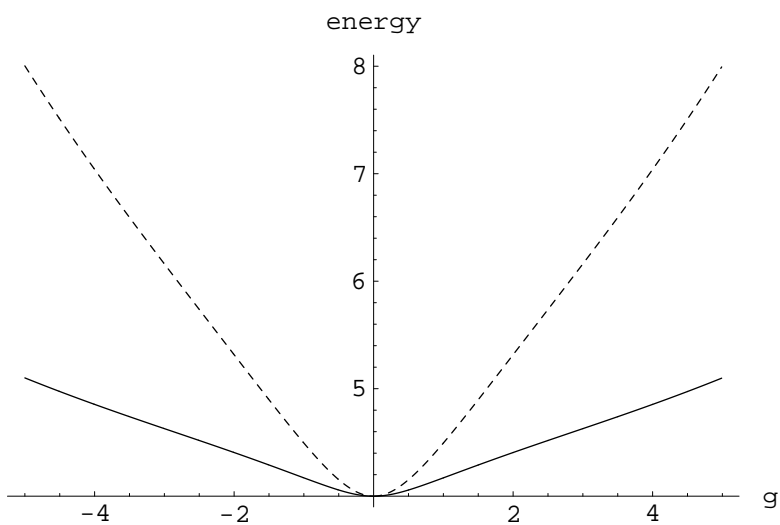}}
\vspace{1cm}
\caption{First two excited energies of the Hamiltonian $H^{(2)}$. The
solid and dashed lines represent the levels whose unperturbed states are $x
e^{-(x^2+y^2)/2}$ and $y e^{-(x^2+y^2)/2}$. The unperturbed energies split
into two distinct levels, both of which lie above the unperturbed level
at $E=4$. The graphs were constructed from the $(9,9)$ Pad\'e of the
perturbation expansion in powers of $g^2$.}
\label{f2}
\end{figure}

The Rayleigh-Schr\"odinger perturbation series for the ground-state energies
$E_0^{(1)}$, $E_0^{(2)}$, and $E_0^{(3)}$ of the Hamiltonians $H^{(1)}$,
$H^{(2)}$, and $H^{(3)}$ have the asymptotic form
\begin{eqnarray}
E_0^{(1)} &\sim& 1+{11\over16}g^2-{465\over256}g^4+{39709\over4096}g^6
-{19250805\over262144}g^8+{2944491879\over4194304}g^{10}+\cdots,\nonumber\\
E_0^{(2)} &\sim& 2+{5\over48}g^2-{223\over6912}g^4+{114407\over4976640}g^6
-{346266143\over14332723200}g^8+{2360833242959\over72236924928000}g^{10}+\cdots,
\nonumber\\
E_0^{(3)} &\sim& 3+{1\over48}g^2-{7\over4608}g^4+{5069\over19906560}g^6-{2441189
\over38220595200}g^8\nonumber\\
&&\qquad\quad+{8034211571\over385263599616000}g^{10}+\cdots
\label{e4}
\end{eqnarray}
in the limit $g\to0$. The $(9,9)$ Pad\'e was constructed from the
once-subtracted form of these series and the results are plotted in
Fig.~\ref{f1}. The first two excited states of $H^{(2)}$ are plotted in 
Fig.~\ref{f2}. Note that the degenerate unperturbed energy level at $E=4$
splits into two levels, each of which is greater than four.

The perturbation coefficients in these series are derived from recursion
relations like those first derived for the anharmonic oscillator \cite{BW}.
These recursion relations are obtained directly from the Schr\"odinger equations
(\ref{e3}). We substitute $\psi_1(x)=e^{-x^2/2}\phi_1(x)$, $\psi_2(x,y)=
e^{-(x^2+y^2)/2}\phi_2(x,y)$ and $\psi_3(x,y,z)=e^{-(x^2+y^2+z^2)/2}\phi_3
(x,y,z)$, where $\phi_{1,2,3}$ are formal power series in $g$; the coefficients
of $g^n$ are polynomials $P_n$ of degree $3n$ in the variables $x$, $y$, $z$.
For example, for the case of $\psi_3$, $P_n=\sum_{j,k,l=0}^n a_{n,j,k,l}x^jy^k
z^l$, and the coefficients $a_{n,j,k,l}$ satisfy
\begin{eqnarray}
a_{n,j,k,l} &=& {1\over2(j+k+l)}\Biggm[
a_{n-1,j-1,k-1,l-1}-2\sum_{p=1}^{n-1}a_{n-p,j,k,l}(a_{p,2,0,0}+a_{p,0,2,0}+
a_{p,0,0,2})\nonumber\\
&&\quad+(j+1)(j+2)a_{n,j+2,k,l}+(k+1)(k+2)a_{n,j,k+2,l}+(l+1)(l+2)a_{n,j,k,l+2}
\Biggm].
\label{e6}
\end{eqnarray}
Once the coefficients $a_{n,j,k,l}$ are known, we can construct the coefficient
of $g^{2n}$ in the expansion for the ground-state energy $E_0^{(3)}$ in
(\ref{e4}) according to
\begin{eqnarray}
E_0^{(3)}\sim 3+2\sum_{n=1}^\infty (a_{2n,2,0,0}+a_{2n,0,2,0}+a_{2n,0,0,2})
(-g^2)^n\quad(g\to0).
\label{e7}
\end{eqnarray}

We are particularly interested in the large-order behavior of the coefficients
in the perturbation expansion because this behavior suggests that the series is
Borel summable and reveals the analytic structure of the energy level as a
function of complex $g^2$. In Ref.~\cite{LO1} it is shown that the large-$n$
behavior of the coefficient of $g^{2n}$ in the expansion of $E_0^{(1)}$ is
\begin{eqnarray}
(-1)^{n+1}{4\over\pi^{3/2}}\left({15\over8}\right)^{n+1/2}\Gamma\left(n+{1\over
2}\right)\left[1-{\rm O}\left({1\over n}\right)\right]\quad(n\to\infty).
\label{e8}
\end{eqnarray}
Therefore, although divergent, the series for $E_0^{(1)}$ is Borel summable
\cite{BO}. Observe that if the factor of $i$ were absent from the Hamiltonian
$H^{(1)}$, then the perturbation coefficients would not alternate in sign and
the perturbation series would not be Borel summable.

A major result reported here is the large-order behavior of the coefficients of
$g^{2n}$ in the series for $E_0^{(2)}$
\begin{eqnarray}
(-1)^{n+1}{72\sqrt{2}\over\pi\sqrt{\cosh({1\over2}\pi\sqrt{23})}}
\left({5\over18}\right)^{n+1/2} \Gamma\left(n+\frac{1}{2}\right)
\left[1-{\rm O}\left(\frac{1}{n}\right)\right] \quad(n\to\infty),
\label{e9}
\end{eqnarray}
and coefficients of $g^{2n}$ in the series for $E_0^{(3)}$
\begin{eqnarray}
(-1)^{n+1}{1152\sqrt{3}\over\sqrt{\pi}\,\cosh({1\over2}\pi\sqrt{23})}
\left({5\over72}\right)^{n+1/2} \Gamma\left(n+\frac{1}{2}\right)
\left[1-{\rm O}\left(\frac{1}{n}\right)\right] \quad(n\to\infty).
\label{e10}
\end{eqnarray}
We have verified these results to extremely high precision by performing a
Richardson extrapolation \cite{BO} of the perturbation coefficients in
(\ref{e4})  divided by these behaviors.

We derive these results by adapting the multidimensional WKB tunneling
techniques in Ref.~\cite{LO3}. We observe that if $g$ is replaced by $ig$ in
$H^{(2)}$ and $H^{(3)}$, then we obtain potentials for which the probability
current in a Gaussian ground state leaks out to infinity. The probability flows
outward along most-probable escape paths (MPEPs).

\begin{figure}[h]
\centering{\epsfig{file=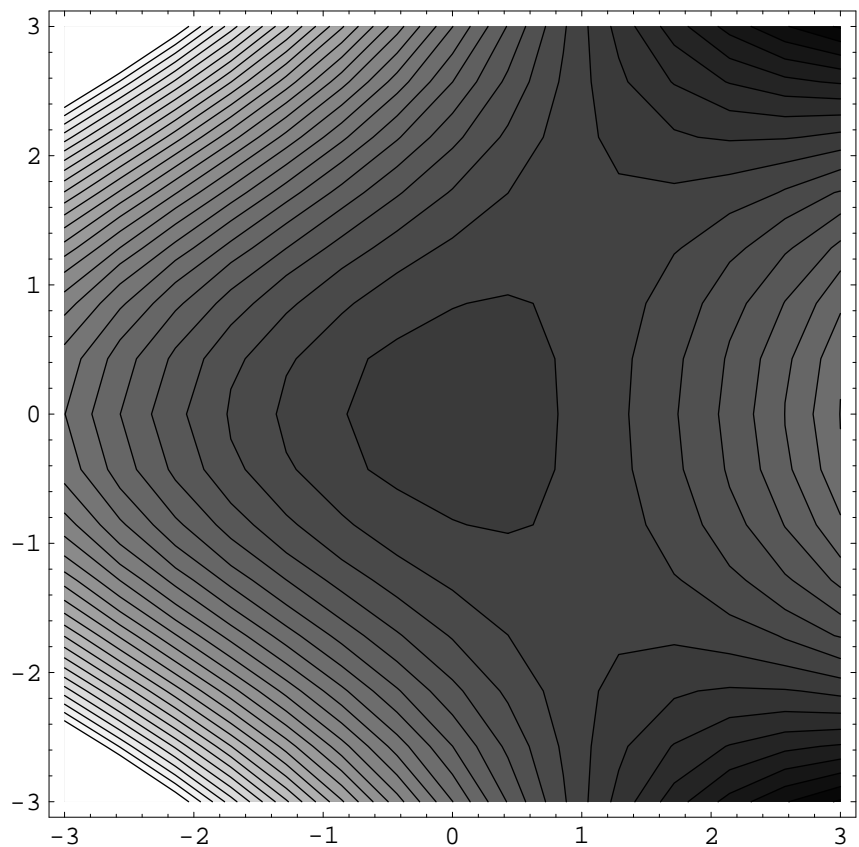}}
\vspace{1cm}
\caption{Contour plot of the potential $x^2+y^2-xy^2$. A Gaussian
probability distribution localized at the origin gradually leaks out to
infinity
preferentially along two channels in the right-half plane. These channels
are called most probable escape paths (MPEPs).}
\label{f3}
\end{figure}

To determine the MPEPs for $H^{(2)}$ we rewrite the potential in polar
coordinates:
\begin{eqnarray}
V(x,y)=x^2+y^2-gxy^2=r^2-gr^3\cos\theta\sin^2\theta,
\label{e11}
\end{eqnarray}
where $x=r\cos\theta$, $y=r\sin\theta$. Letting $\alpha=g\cos\theta\sin^2
\theta$, we calculate $V_r=2r-3\alpha r^2$. Then, setting $V_r=0$ gives the
critical radius $r={2\over3\alpha}$, and at this radius the potential has the
value ${4\over27}\alpha^{-2}$. Thus, $V$ achieves its minimum when $\sin\theta=
\pm\sqrt{2\over3}$. Therefore, the effective radial potential is $V(r)=r^2-{2g
\over3\sqrt{3}}r^3$. We conclude that there are two straight-line MPEPs
symmetrically placed above and below the positive-$x$ axis (see Fig.~\ref{f3}).

{\it Geometrical optics} (ray tracing) is sufficient to reproduce the
gamma-function and exponential behaviors in (\ref{e9}). We simply evaluate the
approximate WKB integral
$$I=-2\int_0^{3\sqrt{3}\over2g} dr\,\sqrt{r^2-{2g\over3\sqrt{3}}r^3},$$
where we have neglected the constant term in the limit of small $g$. We evaluate
the resulting beta-function integral to get the leading exponent in the
tunneling rate: $I=-{18\over5}g^{-2}$. There is a standard
dispersion-integral procedure \cite{LO1,BW} that expresses the large-order
behavior as the $n$th inverse moment of the tunneling rate. This procedure
gives the behavior in (\ref{e9}) apart from an overall multiplicative constant.
This constant can only be determined by performing a {\it physical-optics}
calculation of the tunneling rate.

For this physical-optics calculation we must determine the flux of probability
through a tube centered about the MPEP. We introduce a rotated coordinate system
by
$$x={1\over\sqrt{3}}r-{\sqrt{2}\over\sqrt{3}}t,\quad
y={1\over\sqrt{3}}t+{\sqrt{2}\over\sqrt{3}}r,$$
so that $r$ measures the distance along the MPEP and $t$ is the coordinate
transverse to the MPEP. In terms of these variables, the Schr\"odinger equation
(\ref{e3}) for $\psi_2$ reads
\begin{eqnarray}
-\nabla^2\psi_2(r,t)+\left[r^2+t^2+{g\over3\sqrt{3}}
(-2r^3+3rt^2-\sqrt{3}t^3)-2\right]\psi_2(r,t)=0,
\label{e12}
\end{eqnarray}
where we have replaced $g$ by $ig$. We may drop the $t^3$ term because
$gt^3<<t^2$ for small $g$.

Next, we separate the radial dependence from the transverse dependence by
writing $\psi_2(r,t)=W(r)\phi(r,t)$. The function $W(r)$, which expresses the
radial dependence, satisfies the differential equation $W''(r)=\left(r^2-{2g
\over3\sqrt{3}}r^3-1\right)W(r)$. The WKB approximation to the decaying solution
to this equation is
\begin{eqnarray}
W(r)={e^{-1/4}\exp\left(-\int_1^r ds\,\sqrt{s^2-{2g\over3\sqrt{3}}s^3-1}\right)
\over\sqrt{2}\,\left(r^2-{2g\over3\sqrt{3}}r^3-1\right)^{1/4}},
\label{e13}
\end{eqnarray}
where the numerical factors are included in anticipation of asymptotic matching.
The equation for $\phi(r,t)$ is $-2{W_r\over W}\phi_r-\phi_{rr}-\phi_{tt}+(t^2+
{g\over\sqrt{3}}rt^2-1)\phi=0$. Note that $W_r/W\sim-r$ for small $g$. The
change of variable $v=\sqrt{1-{2g\over3\sqrt{3}}r}$ yields a parabolic equation
for $\phi$:
\begin{eqnarray}
(v^2-1)\phi_v-\phi_{tt}+[t^2+3t^2(1-v^2)/2 -1]\phi=0,
\label{e14}
\end{eqnarray}
where we neglect the small term of order $g^2\phi_{vv}$. The solution to this
equation has the form of a Gaussian that expresses the thickness of the stream
of probability current that flows outward along the MPEP:
\begin{eqnarray}
\phi(v,t)=A(v)e^{-t^2f(v)/2},
\label{e15}
\end{eqnarray}
where the function $f(v)$ satisfies the Riccati equation
\begin{eqnarray}
(1-v^2)f'(v)-2f^2(v)+5-3v^2=0
\label{e16}
\end{eqnarray}
and $A(v)$ satisfies the transport equation
\begin{eqnarray}
(1-v^2)A'(v)-f(v)A(v)+A(v)=0.
\label{e17}
\end{eqnarray}

To solve the Riccati equation (\ref{e16}) we substitute $f(v)=-{1\over2}(1-v^2)
h'(v)/h(v)$ and convert it to the second-order linear equation
\begin{eqnarray}
(1-v^2)h''(v)-2vh'(v)+\left(-6-{4\over1-v^2}\right)h(v)=0,
\label{e18}
\end{eqnarray}
which we recognize as the Legendre differential equation \cite{BMP}.

To find the initial conditions on $f(v)$ and $A(v)$, we match $\phi(v,t)$ to the
wave-function solution to the Schr\"odinger equation (\ref{e12}) in the inner
region where $r$ and $t$ are of order $1$. In this region the wave function is a
Gaussian: $\phi(v,t)=e^{-(r^2+t^2)/2}$. By construction, for small $r$, $W(r)
\sim e^{-r^2/2}$. Thus, we obtain the initial conditions $A(1)=1$ and $f(1)=1$.
Hence, the solution to (\ref{e18}) is the Legendre function $h(v)=P_\nu^{-2}
(v)$, where $\nu(\nu+1)=-6$.

Our objective now is to find the flux of probability at the distant turning
point \cite{LO3}. At this turning point the radial component of the probability
current is
\begin{eqnarray}
J=2{1\over2\sqrt{e}}e^{-t^2f(0)}A^2(0)\exp\left(-2\int_1^{3\sqrt{3}\over2g}ds\,
\sqrt{s^2-{2g\over3\sqrt{3}}s^3-1}\right),
\label{e20}
\end{eqnarray}
where we have included a factor of 2 because there are two channels. We
integrate in the transverse direction to get the total flux of probability
current $\int_{-\infty}^{\infty} dt\, e^{-t^2f(0)}=\sqrt{\pi\over f(0)}$ and we
evaluate the integral in the exponent to obtain
$$2\int_1^{3\sqrt{3}\over2g} ds\,\sqrt{s^2-{2g\over3\sqrt{3}}s^3-1}\sim
{18\over5g^2}-\ln(12\sqrt{3})+\ln(g)-{1\over2}\quad(g\to0).$$
Thus, the total outward flux of probability is ${12\sqrt{3}\over g}A^2(0)\sqrt{
\pi\over f(0)}e^{-18/(5g^2)}$. Substituting this result into the dispersion
integral, we obtain the following formula for the large-order behavior of the
coefficients in the perturbation series for the ground-state energy:
\begin{eqnarray}
(-1)^{n+1}{12A^2(0)\sqrt{3}\over\pi\sqrt{\pi f(0)}} \left({5\over18}\right)^{
n+1/2} \Gamma\left(n+\frac{1}{2}\right) \left[1-{\rm O}\left(\frac{1}{n}\right)
\right] \quad(n\to\infty).
\label{e22}
\end{eqnarray}

It remains to find the numbers $f(0)$ and $A(0)$. Using the
hypergeometric-function representation for the Legendre function \cite{BMP}, we
obtain
$$f(0)=-\tan(\pi\nu/2)\,{\Gamma(\nu/2)\Gamma(2+\nu/2)\over\Gamma(-1/2+\nu/2)
\Gamma(3/2+\nu/2)}$$
and
$$A(0)=\pi^{-1/4}\,\sqrt{2\Gamma(2+\nu/2)\Gamma(3/2-\nu/2)}.$$
Thus, $A^2(0)/\sqrt{f(0)}=\sqrt{24\pi/\cosh(\pi\sqrt{23}/2)}$ and we have
derived the result in (\ref{e9}).

To obtain the formula (\ref{e10}) we follow the same procedure. In this case
there are four radial MPEPs and there are two transverse variables. We begin
by introducing a change of coordinates in the Schr\"odinger equation (\ref{e3})
for $\psi_3(x,y,z)$:
$$x={1\over\sqrt{3}}r+{1\over\sqrt{6}}s-{1\over\sqrt{2}}t,\quad
y={1\over\sqrt{3}}r+{1\over\sqrt{6}}s+{1\over\sqrt{2}}t,\quad
z={1\over\sqrt{3}}r-{2\over\sqrt{6}}s.$$
The remainder of the calculation is identical to that summarized above for
$\psi_2(x,y)$.

\begin{figure}[h]
\centering{\epsfig{file=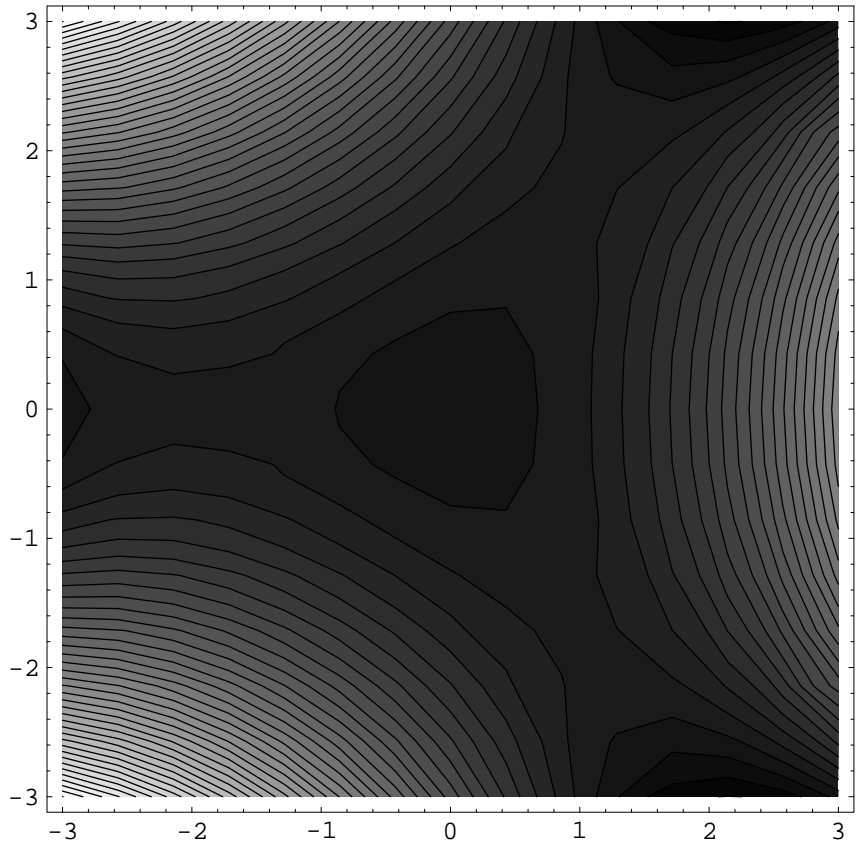}}
\vspace{1cm}
\caption{Contour plot of the H\'enon-Heiles potential
$x^2+y^2-xy^2+{1\over3}x^3$. A Gaussian probability distribution localized
at the origin tunnels out
to infinity preferentially along three MPEPs, two in the right-half plane
and
one along the negative-real axis. Remarkably, the contribution from all
three
MPEPs to the large-order behavior of perturbation theory is of the same
magnitude.}
\label{f4}
\end{figure}

We conclude by noting that the complex H\'enon-Heiles potential
\begin{equation}
V^{HH}=x^2+y^2+i g (xy^2-{1\over3}x^3)
\label{e23}
\end{equation}
has a Borel summable Rayleigh-Schr\"odinger perturbation series for the
ground-state energy, with the leading growth of the coefficients given by
\begin{equation}
C(-1)^{n+1}\,\left({5\over24}\right)^{n+1/2}\,\Gamma\left(n+{1\over2}\right)\,
\left[1+O\left({1\over n}\right)\right] \quad (n\to\infty),
\label{e24}
\end{equation}
where $C$ is a constant. A simple geometric-optics calculation (with $g$
replaced by $ig$) confirms this leading growth rate. There are three MPEPs, a
pair of MPEPs similar to those encountered in our analysis of $H^{(2)}$ (now at
angles $\pm{\pi\over3}$ from the positive-$x$ axis), and a third MPEP along the
negative-$x$ axis, which is like that for $H^{(1)}$ (see Fig.~\ref{f4}).
Remarkably, these two different types of MPEPs produce exactly the same leading
contribution to (\ref{e24}). This fact depends crucially on having the
appropriate combinatorial factors in (\ref{e23}).

Finally, we remark that the quantum field theoretic generalizations of the
Hamiltonians studied here, particularly $H^{(2)}$, may be viewed as theories of
scalar electrodynamics. It would be interesting to study such issues as bound
states and Schwinger-Dyson equations in such theories.

\section*{ACKNOWLEDGMENTS}
\label{s1}
M\d{S} is grateful to the Physics Department at Washington University for their
hospitality during his sabbatical. This work was supported by the
U.S.~Department of Energy.

%\begin{figure}
%\vspace{4.5in}
%\special{psfile=fig1.eps angle=0 hoffset=-65 voffset=15 hscale=120
%vscale=120}
%\caption{Ground-state energies of the Hamiltonians $H^{(1)}$ (solid line),
%$H^{(2)}$ (long-dashed line), and $H^{(3)}$ (short-dashed line), as
%functions of the coupling constant $g$. Note that the energy levels are
%real and positive. The graphs were obtained from the $(9,9)$ Pad\'e
%constructed from the once-subtracted perturbation series for these energy
%levels.}
%\label{f1}
%\end{figure}

\newpage
\center{FIGURE 2}

%\begin{figure}
%\vspace{4.5in}
%\special{psfile=fig2.eps angle=0 hoffset=-65 voffset=15 hscale=120
%vscale=120}
%\caption{First two excited energies of the Hamiltonian $H^{(2)}$. The solid
%and dashed lines represent the levels whose unperturbed states are $x
%e^{-(x^2+y^2)/2}$ and $y e^{-(x^2+y^2)/2}$. The unperturbed energies split
%into two distinct levels, both of which lie above the unperturbed level
%at $E=4$. The graphs were constructed from the $(9,9)$ Pad\'e of the
%perturbation expansion in powers of $g^2$.}
%\label{f2}
%\end{figure}

\newpage
\center{FIGURE 3}

%\begin{figure}
%\vspace{4.5in}
%\special{psfile=fig3.eps angle=0 hoffset=-25 voffset=10 hscale=120
%vscale=120}
%\caption{Contour plot of the potential $x^2+y^2-xy^2$. A Gaussian
%probability distribution localized at the origin gradually leaks out to
%infinity preferentially along two channels in the right-half plane. These
%channels are called most probable escape paths (MPEPs).}
%\label{f3}
%\end{figure}

\newpage
\center{FIGURE 4}

%\begin{figure}
%\vspace{4.5in}
%\special{psfile=fig4.eps angle=0 hoffset=-25 voffset=10 hscale=120
%vscale=120}
%\caption{Contour plot of the H\'enon-Heiles potential
%$x^2+y^2-xy^2+{1\over3} x^3$. A Gaussian probability distribution localized
%at the origin tunnels out to infinity preferentially along three MPEPs, two
%in the right-half plane and one along the negative-real axis. Remarkably,
%the contribution from all three MPEPs to the large-order behavior of
%perturbation theory is of the same magnitude.}
%\label{f4}
%\end{figure}

\end{document}